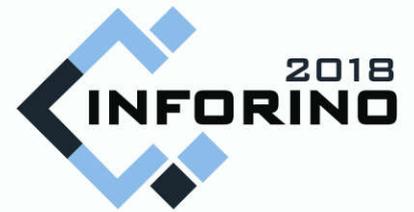

IV International Conference on

# INFORMATION TECHNOLOGIES IN ENGINEERING EDUCATION

23-26 October, 2018

Moscow, Russia

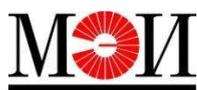
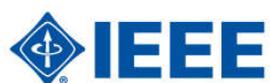
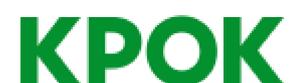

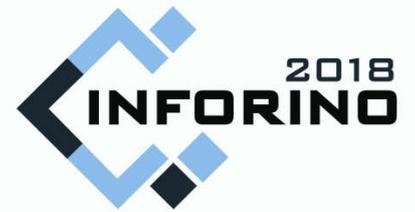

SECTION 2

# Software and Information Support for Engineering Education

# Quantitative Assessment of Solution Innovation in Engineering Education

V.K. Ivanov, A.G. Glebova, I.V. Obrazthov
Tver State Technical University
Tver, Russia
mtivk@mail.ru, nauka_rf@mail.ru, sunspire@list.ru

*Abstract*—The article discusses the quantitative assessment approach to the innovation of engineering system components. The validity of the approach is based on the expert appraisal of the university's electronic information educational environment components and the measurement of engineering solution innovation in engineering education. The implementation of batch processing of object innovation assessments is justified and described.

I. INTRODUCTION

According to the report [1] about 66% of chief academic officers believe online education to be critical to their long-term strategy. Globally, there is a trend: the more the higher school enrolment and the higher the level of education programs proposed (Master, Doctorate studies, training and retraining programs) are, the more often the higher school uses e-learning.

One of the present trends in university engineering training is the transition to blended learning whose development is impossible without innovation educational solutions. Especially since the Federal State Educational Standards (FSES) of higher education for all training levels and specialties determine the requirements to educational program implementation providing [2, Section VII] the obligatory application of electronic information educational environment (EIEE) being the infrastructure base of a university educational process.

On the other hand, students – future engineers – learn to research, design, develop and introduce engineering systems of different complexity and novelty. It is obvious that information technologies should be used extensively in assessing their engineering solution innovation.

The article examines the quantitative assessment approach to the innovation of engineering system components. The authors show its validity with the expert appraisals of the university's EIEE components and the measurement of engineering solution innovation in engineering education. They also justify and describe the implementation of batch processing of object innovation assessments.

II. APPROACH TO INNOVATION ASSESSMENT

The analysis of [3, 4, 5, 6], devoted to different issues of innovation development, shows that 'innovation' always includes such connotations as new, scientific, efficiency-increasing, profit-making. Hence, innovation will be used to mean a product (object) having the set of properties determining its technological novelty, relevance, and implementability.

Technological novelty means significant improvements, a new way of using or granting a product (a system component) or a technology. Novelty subjects are potential users or producers themselves.

Relevance of a product means that it is required by a potential user feeling and requesting it as a necessity. In other words, relevance is the capacity of a product claiming innovation to meet a potential user's requirements. It can be determined with the appraisal of potential users' requests and / or the appraisal of using functional counterparts.

Implementability of a product determines the technological validity, physical feasibility and capacity of a product to be integrated in the system in order to have desired effects.

An important conclusion which emerges from these definitions is that innovation has a system property, i.e. a product or service innovation must be considered within its system relations, the degree of the component influence on a subject and external objects. At the same time the dynamics of indicators (increase, stability, and decrease) should be taken into account.

It should be mentioned that some product or service properties defining innovation in general are beyond the scope of the present article and not discussed. They are the availability of a patent (or patentability), cost-effectiveness, and judicial admissibility.

III. EXPERT APPRAISAL OF SYSTEM COMPONENT INNOVATION

We shall consider the approach to determining the innovation of technical (information) system components with expert appraisals. The example of the system is the University's EIEE. According to Para 3, Article 16 of Federal Law No. 273-FZ of December 29, 2012, (revised March 07, 2018) On Education in the Russian Federation electronic information educational environment (EIEE) includes electronic information resources, electronic educational resources, information and telecommunication technologies, appropriate technological facilities and provides students with the possibility of mastering educational programs fully regardless of their location. Order of the RF Ministry of Education and Science No. 816 of August 23, 2017 On Approval of the Procedure for Applying E-Learning and Distance Learning Technologies by Educational Organizations When Delivering Study Programs provides the opportunity for educational organizations to implement training curricula or their parts with e-learning and distance learning technologies on the basis of its EIEE and (or) other organizations' resources.

EIEE is an integrated information system providing the interoperability of different components by sharing online and reference data as well as program and user interface. The architecture, the implementation platform, and the exhaustive composition of functional components of Tver State Technical University (TvSTU) EIEE have been discussed repeatedly (see, e.g., [7]).

We can identify the following EIEE components being considered as innovative mechanisms: e-learning environment, a data base of teaching and learning materials, a student electronic portfolio, a monitoring subsystem of training process and results, an electronic library system, a videoconference and webinar server, virtual labs and workshops. Each EIEE components can be characterized from the point of its innovation.

Expert appraisals of innovation indicators (novelty, relevance, implementability) should be carried out by the participants in educational process (students, teachers, management). Basic ('low', 'medium', 'high', 'irrelevant') and auxiliary ('stable', 'increasing'. 'decreasing') estimation scales are used.

EIEE innovation level is assessed with an electronic enquiry of experts from several groups followed by statistic processing of the results obtained in accordance with Evidence Theory which application is specified by the inaccuracy (interval character) of initial expert appraisals. Further the article describes the implementation of the coprocessing algorithm of expert appraisals and / or innovation indicator measurements. The implementation is based on Evidence Theory (Section V).

If expert appraisals can be assigned numerical values and weight coefficients showing the indicator relevance to others, the overall innovation indicator of EIEE component, a group of EIEE components, and EIEE on the whole will be the following:

$$In = \sum_{k=1}^{K}\sum_{j=1}^{J}\sum_{i=1}^{I} w_{i,j,k} V_{i,j,k} \quad (1)$$

where *In* is an integral innovation index of a group of EIEE components; *K* is a number of components; *J* is a number of EIEE user groups (*J*=3 if user groups 'students', 'teachers', and 'management' are considered); *I* is a number of innovation indicators (*I*=3 if innovation indicators 'novelty', 'relevance', and 'implementability' are considered); $w_{i,j,k}$ is a weight coefficient for *i*-th innovation indicator, *j*-th group of users, and *k*-th EIEE component, $\sum_{k=1}^{K} w_k = \sum_{j=1}^{J} w_j = \sum_{i=1}^{I} w_i = 1$; $V_{i,j,k}$ is a numerical value for *i*-th innovation indicator, *j*-th group of users, and *k*-th EIEE component (basic and auxiliary scale values are used).

The use of the EIEE component innovation index *In* is expedient in the following cases:

- Assessing the scientific and technological level of the components and EIEE on the whole.
- Planning EIEE innovation development (not just improving some indicators but changing relations within the system).
- Generating new project solutions.
- Assessing the degree of EIEE goal implementation and use.

IV. ASSESSMENT OF ENGINEERING SOLUTION INNOVATION IN ENGINEERING EDUCATION

The approach to determining product or technology innovation is also based on direct measurements of indicator values which may be used as quantitative assessments of some innovation properties. The assessment of engineering solution innovation in engineering education is particularly relevant, given the development strategy of the region industries worked out in accordance with the Russian Federation Government requirements [8]. We suggest to discuss the use of the approach in education as we believe that it will help assess properly the results of researching, designing, and developing complex engineering systems and contribute to better understanding of what engineering system innovation is.

Consider the idea of the object novelty measurement. Imagine that some data storage contains various and poorly structured, as a rule, text information about this and other objects. They may be special-purpose data bases and / or the Internet resources on the whole. The information stored may be classified by time intervals. There is a retrieval pattern for the object. It is a set of keywords and notions defining the object accurately in the following areas: the application (operation) mechanism or construction, the application (operation) result or conditions, crucial characteristics (properties, material, composition). Also there is a marker or a term determining the application area or object actions.

The novelty factor of object $\mathbb{N}$ is determined as follows:

$$\mathbb{N} = 1 - \frac{\frac{1}{Q}\sum_{q=1}^{Q} N_q}{N_m} \quad (2)$$

where $N_q$ is the number of storage units (files, records) found in the storage used in executing the *q*–th retrieval query generated from a retrieval pattern; *Q* is the number of retrieval queries generated from a retrieval pattern; $N_m$ is the number of storage units found in the storage used with a marker. The value of $\mathbb{N}$ is ranged to [0;1].

Our study of engineering solution parameters by the scheme above proves its validity.

Figures 1 and 2 show the novelty assessment results of TOP-10 inventions of 2017 from the list prepared by Rospatent (the Federal Service for Intellectual Property) experts [9] and ten randomly selected inventions registered in 2017. The novelty factor $\mathbb{N}$ for each invention and the average value $\bar{\mathbb{N}}$ for the group were estimated. Retrieval patterns were prepared by hand. The open patent data base http://www1.fips.ru was used. The values $\bar{\mathbb{N}}$ obtained with experiments and estimations show that the TOP-10 invention novelty is higher than the novelty of ten randomly selected inventions. This was quite expected.

The analysis of engineering solution novelty assessed for a certain period also brings some interesting conclusions. Figure 3 shows the results of the experiments conducted to determine the novelty factor $\mathbb{N}$ of an engineering solution described in the patent Method of Chocolate Mass Production. The patent data base http://www1.fips.ru is used. The novelty factor $\mathbb{N}$ for the patent under study is determined separately by data base subsets corresponding to the years of patent registration (a ten-year period from 2008 to 2017 is used).

It is not difficult to see that linear, power, and exponential approximations and the corresponding trend lines prove the hypothesis of reducing the object novelty in time. It is logical that the diffusion and use of a new object idea make the object conventional and commonplace. The more information on the object the community and various data bases have, the lower level of the object novelty the estimations show.

The polynomial approximation shows the novelty cycle (and, hence, the innovation cycle) of the analyzed object that requires the check of the hypothesis of five-year innovation cycles in food industry. The development cycle is known to be characteristic of all systems. The widely-known concepts such as a sequence of recurring cycles or long waves describe the development of a society in general and that of an economy in particular [10, 11]. Certainly, it only concerns the hypothesis generation which needs to be followed up.

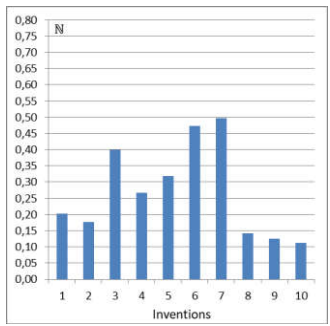

Fig. 1. The novelty factor N for TOP-10 inventions of 2017, $\bar{\mathbb{N}} = 0,27$.

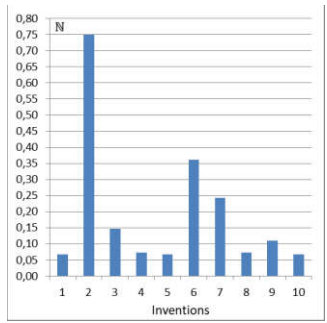

Fig. 2. The novelty factor N for randomly selected inventions of 2017, $\bar{\mathbb{N}} = 0,20$.

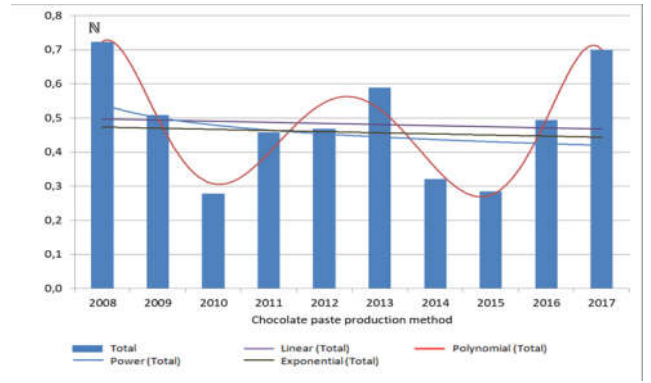

Fig. 3. The novelty factor for the patent Method of Chocolate Mass Production, 2008-2017.

## V. JUSTIFICATION AND IMPLEMENTATION OF BATCH PROCESSING OF OBJECT INNOVATION ASSESSMENTS

Actually, expert appraisals (1) have, as a rule, value ranges. Mathematical modeling and processing of inaccurate (interval) expert appraisals, measurements or observations may be carried out with Dempster–Shafer theory (Evidence Theory) [12]. In our case, the integral innovation index is assumed to be *P*(*In*) as a probability of getting *In* into some range while being

$$Bel(In) < P(In) < Pl(In) \qquad (3)$$

where *Bel*(*In*) is a function of belief in the expert appraisal (evidence); *Pl*(*In*) is a function of plausibility of an expert appraisal.

Several evidences are combined in accordance with the formal combination rule (Dempster's rule).

It should be mentioned that some indicators can be results of relatively accurate estimations instead of expert appraisals. For example, the level of EIEE component use (relevance) can be estimated as an indicator of EIEE effectiveness as follows:

$$Demand = LF_d / LF \qquad (4)$$

where $LF_d$ is the number of staff d labour functions which are performed with EIEE components, *LF* is the total number of labour functions.

In addition, implementability can be estimated with the counted-up *Pr* value which is the number of unsolved problems of using EIEE by the staff performing their labour functions.

Basic mathematical estimations in accordance with Dempster–Shafer theory of evidences (based on the interval assessments) involves the work with big volumes of data especially if we speak about the assessment of a complex multicomponent system whose innovation is assessed by many indicators. In this case software should be used to structure data and provide the user with the convenient form of basic mathematical operations for estimating the values of belief and plausibility functions of one or another evidence. The result is an array of combined estimations for each indicator of all the system components under study and an integral assessment of the system in the form of numerical assessment interval. The derived integral assessments of a multicomponent system show its innovation objective level. The present approach allows the assessment time to be reduced thanks to the optimal structuring of experts' inquiry results, observations or measurements with grouping and sorting the data by many categories, with basic estimations being automated.

Categorized parameters (components, indicators, expert groups and assessments) are basic initial data to create a program system for the batch processing of innovation level assessments. The following algorithm (based on EIEE) is proposed:

1. Collect initial data (lists of components, indicators, expert groups and assessments).

2. Input data by categories: 'components', 'indicators', 'expert groups', 'estimation scale', 'inquiry results'.

3. Create evidence tables – sort out inquiry results and combine equal assessments.

4. Estimate basic evidence probabilities:

$$m(A_i) = \frac{C_i}{N_i} \qquad (5)$$

where $m$ is a basic probability (mass) of the $i$-th evidence with a numerical interval $A$, $C_i$ is the number of experts in an expert group being the source of the evidence, $N_i$ is the total number of experts in an expert group.

5. Combine evidences from different expert groups by components and indicators:

$$m_{1...j}(A) = \frac{\sum_{A_{1,1} \cap ... \cap A_{p,j}} \prod_{j=1}^{n} m_j(A_{p,j})}{1-K} \qquad (6)$$

where $n$ is the number of independent evidence sources (expert groups), $A_{1,1} \cap ... \cap A_{p,j} = A$ is an intersection of numerical evidence intervals from independent sources, $m_j(A_{p,j})$ are basic probabilities of evidences from independent sources which numerical intervals intersect, $K$ is a normalization factor of conflict (nonintersecting evidences) estimated as follows:

$$K = \sum_{A_{1,1} \cap ... \cap A_{p,j} = \varnothing} \prod_{j=1}^{n} m_j(A_{p,j}) \qquad (7)$$

where $A_{1,1} \cap ... \cap A_{p,j} = \varnothing$ is a non-intersection condition of numerical evidence intervals from independent sources.

6. Determine numerical boundaries of the expected interval:

$$\underline{EX} = \sum_{i=1}^{n} m(A_i) \cdot \inf A_i \qquad (8)$$

$$\overline{EX} = \sum_{i=1}^{n} m(A_i) \cdot \sup A_i \qquad (9)$$

7. Determine the value of belief and plausibility functions of the evidences combined:

$$Bel(A) = \sum_{A_i \in A} m(A_i) \qquad (10)$$

$$Pl(A) = \sum_{A_i \cap A \neq \varnothing} m(A_i) \qquad (11)$$

8. Create the table of integral assessments by components and / or indicators.

It should be mentioned that the combination of evidences obtained from different sources (expert groups) is a crucial step of the algorithm. The goal of evidence combination is to combine evidences whose numerical intervals intersect and to take into account conflict evidences whose numerical intervals do not intersect. The combination is done with Dempster's formal rule recursively by source pairs. One iteration checks the conditions of numerical interval intersections and estimates the total basic probabilities with equations (6) and (7), i.e. two evidence sources form one conditional source whose evidences are combined with the following actual source. Numerical boundaries of the expected interval (the boundaries of mathematical expectation) are estimated for all the combined evidences with equations (8) and (9). The functions of belief (10) and plausibility (11) are the lower and upper boundaries of evidence probability respectively. The maximal value of belief and plausibility shows the most probable assessment which is entered in the integral assessment table.

The algorithm described is implemented as a web-based application with GUI (Figure 4). The application is designed in SpiderBasic whose compiler generates the optimized code of JavaScript requiring a browser supporting HTML5.

In addition to the assessment of engineering solution innovation we propose the introduction of the application in different engineering areas where data sources can have a posteriori information such as electronic pickups or digital measuring devices. Another field of development is the support of mathematical methods being alternative to Evidence Theory.

Fig. 4. GUI of the expert appraisal of a multicomponent system based on Dempster–Shafer theory

## VI. CONCLUSIONS

The innovation practice of developed countries in education means the implementation of appropriate programs as a part of public policy. Moreover, the emphasis is made on the strengthening cooperation of universities and business entities (see, e.g., [13]). In this case, the works providing the educational technology innovation comparable to the best world achievements are definitely necessary. On the other hand, the assessment of engineering solution innovation being done in the educational process of an engineering higher school is no less critical. It should be mentioned that in 2017 TvSTU is included in the list of Federal innovation sites for 2018-2023. We hope that the present research will help intensify the work on formalizing the project component innovation, analyzing and assessing the modernization level of project solutions, specifying the estimations of operating benefits, and, further, using the obtained solutions and experience in domestic and overseas educational practice.

ACKNOWLEDGMENT

This work was supported by RFBR (Project No. 18-07-00358).